\documentclass[aps, pra, twocolumn,showpacs,preprintnumbers,amsmath,amssymb]{revtex4}

\newcommand{\ket}[1]{\mbox{$ | #1 \rangle $}}

\usepackage[dvips]{graphicx}

\begin{document}

\preprint{}

\title{Efficient multiparty quantum secret sharing of secure direct communication}

\author{Jian Wang}

 \email{jwang@nudt.edu.cn}

\affiliation{School of Electronic Science and Engineering,
\\National University of Defense Technology, Changsha, 410073, China }
\author{Quan Zhang}
\affiliation{School of Electronic Science and Engineering,
\\National University of Defense Technology, Changsha, 410073, China }
\author{Chao-jing Tang}
\affiliation{School of Electronic Science and Engineering,
\\National University of Defense Technology, Changsha, 410073, China }



\begin{abstract}
In this paper, we present an ($n, n$) threshold quantum secret
sharing scheme of secure direct communication using
Greenberger-Horne-Zeilinger state. The present scheme is efficient
in that all the Greenberger-Horne-Zeilinger states used in the
quantum secret sharing scheme are used to generate shared secret
messages except those chosen for checking eavesdropper. In our
scheme, the measuring basis of communication parties is invariable
and the classical information used to check eavesdropping needs only
the results of measurements of the communication parties. Another
nice feature of our scheme is that the sender transmit her secret
messages to the receivers directly and the receivers recover the
sender's secret by combining their results, different from the QSS
scheme whose object is essentially to allow a sender to establish a
shared key with the receivers. This feature of our scheme is similar
to that of quantum secret direct communication.
\end{abstract}

\maketitle

\section{Introduction}
%
%
Quantum key distribution (QKD) is one of the most promising
applications of quantum information science. The goal of QKD is to
allow two legitimate parties, Alice and Bob, to generate a secret
key over a long distance, in the presence of an eavesdropper, Eve,
who interferes with the signals. QKD has progressed quickly since
Benneett and Brassard designed the original QKD protocol\cite{bb84}.
Recently, a novel concept, quantum secure direct communication
(QSDC) has been proposed and pursued\cite{beige,Bostrom,Deng}.
Different from QKD, QSDC's object is to transmit the secret messages
directly without first establishing a key to encrypt them. QSDC can
be used in some special environments which has been shown by
Bostr\"{o}em and Deng et al.\cite{Bostrom,Deng}.

Quantum secret sharing (QSS)\cite{hbb99,kki99} is another important
application of quantum mechanics. The basic idea of secret sharing
in the simplest case is that the sender Alice splits the secret
message into two shares and distributes them to two receivers Bob
and Charlie separately, such that only the two receivers collaborate
can they reconstruct the secret message. In a more general setting,
a $(m, n)$ threshold scheme, the secret message is split into $n$
shares, such that any $m$ of those shares can be used to reconstruct
it. QSS is the generalization of classical secret sharing and can
share both classical and quantum messages. QSS is likely to play a
key role in protecting secret quantum information, e.g., in secure
operations of distributed quantum computation, sharing
difficult-to-construct ancillary states and joint sharing of quantum
money, etc. Many researches have been carried out in both
theoretical and experimental aspects after the pioneering QSS scheme
proposed by Hillery, Buz\v{e}k and Berthiaume in 1999 (hereafter
called HBB99)\cite{hbb99}. The HBB99 scheme is based on a
three-particle entangled Greenberger-Horne-Zeilinger (GHZ) state.
Karlsson, Koashi and Imoto\cite{kki99} proposed a QSS scheme using
two-particle Bell states. Guo-Ping Guo and Guang-Can Guo\cite{gg03}
presented a QSS scheme where only product states are employed.
Zhan-jun Zhang, Yong Li and Zhong-xiao Man\cite{zlm05} proposed a
QSS scheme using single photons.

The efficiency is one of the important parameters of quantum
protocol. In the HBB99 scheme, only half of the GHZ states can be
used to generate shared secret messages, because the communication
parties, Alice, Bob, and Charlie need to choose randomly one
measuring basis from either the $X$-basis or the $Y$-basis,
respectively. Recently, Li Xiao, Gui Lu Long, Fu Guo Deng, and Jian
Wei Pan\cite{xldp04} generalized the HBB99 scheme into arbitrary
multiparties and improved the efficiency of the QSS scheme by two
techniques from quantum key distribution (hereafter called XLDP04).
One of the two techniques is that all the participants choose their
measuring basis asymmetrically\cite{lca00}, another one is that all
the participants choose their measuring basis according to a control
key\cite{hkh98}. However, the flaws of the above two techniques are
obvious. The scheme using the first technique needs a refined data
analysis and the scheme using the other one requires the
communication parties share a common key.

In this paper, we present an efficient multiparty QSS scheme using
GHZ state and its transformation. It has the high intrinsic
efficiency as all the states are used for secret sharing except that
chosen for eavesdropping check. It is not necessarily for the
communication parties to chose measuring basis for the measuring
basis is invariable in our scheme. In the eavesdropping check of the
scheme, the communication parties need only to announce their
results of measurements. Our scheme provides higher efficiency than
that of the XLDP04 scheme, for the intrinsic shortage of their
scheme. Moreover, the sender, Alice can transmit her secret messages
directly to the receivers, which is similar to QSDC. The receivers
can only recover the secure direct transmitted messages by combining
their results, after Alice announces her results. Therefore the
present scheme may also be called multiparty quantum secure direct
communication scheme.

This paper is organized as follows. In Sec.\ref{3-protocol}, we
describe the three-party QSS scheme. In Sec.\ref{security}, we
discuss the security of the present scheme. In Sec.\ref{n-protocol},
we generalize the three-party QSS scheme to $(n, n)$ threshold
scheme. Finally, we give a summary in Sec.\ref{conclusion}.

\section{Three-party QSS scheme}
\label{3-protocol}
In the three-party QSS scheme, we suppose the sender Alice wants to
send a secret message to two receivers, say Bob and charlie, so that
none of the receivers can recover the messages on his own. The basic
idea of the scheme originates from quantum
teleportation\cite{bbc93}. Alice entangles his encoded secret
message state with a prepared three-particle entangled state. She
then performs controlled-NOT (CNOT) operation and Hadamard
transformation, which is similar to the method used in quantum
teleportation. Different from quantum teleportation, the receivers
measure their particles in a fixed measuring basis instead of
performing unitary operation to recover the sending qubit with
Alice's classical message. The three-party QSS scheme is as follows:

(1) Alice prepares an ordered $N$ three-particle states. Each of
three-particle states is randomly in the state
\begin{eqnarray}
\ket{\psi_1}=\frac{1}{\sqrt{2}}(\ket{000}+\ket{111})_{ABC},
\end{eqnarray}
\begin{eqnarray}
\ket{\psi_2}=\frac{1}{\sqrt{2}}(\ket{0+0}+\ket{1-1})_{ABC},
\end{eqnarray}
\begin{eqnarray}
\ket{\psi_3}=\frac{1}{\sqrt{2}}(\ket{00+}+\ket{11-})_{ABC},
\end{eqnarray}
\begin{eqnarray}
\ket{\psi_4}=\frac{1}{\sqrt{2}}(\ket{0++}+\ket{1--})_{ABC},
\end{eqnarray}
where $\ket{+}=\frac{1}{\sqrt{2}}(\ket{0}+\ket{1})$,
$\ket{-}=\frac{1}{\sqrt{2}}(\ket{0}-\ket{1})$. We denotes the
ordered $N$ three-particle qubits with
\{[P$_1(A)$,P$_1(B)$,P$_1(C)$], [P$_2(A)$,P$_2(B)$,P$_2(C)$],
$\cdots$, [P$_N(A)$,P$_N(B)$,P$_N(C)$]\}, where the subscript
indicates the order of each three-particle in the sequence, and $A$,
$B$, $C$ represents the three particles of each state, respectively.
Alice takes one particle from each state to form an ordered partner
particle sequence [P$_1(A)$, P$_2(A)$,$\cdots$, P$_N(A)$], called
$A$ sequence. The remaining partner particles compose $B$ sequence,
[P$_1(B)$, P$_2(B)$,$\cdots$, P$_N(B)$] and $C$ sequence, [P$_1(C)$,
P$_2(C)$,$\cdots$, P$_N(C)$]. Alice sends $B$ sequence and $C$
sequence to each Bob and Charlie. Bob and Charlie then inform Alice
that they have received $N$ particles, respectively.

(2) After hearing from Bob and Charlie, Alice announces publicly
each of three-particle states she prepared. If the state of
[P$_i(A)$,P$_i(B)$,P$_i(C)$] is $\ket{\psi_2}$ ($\ket{\psi_3}$), Bob
(Charlie) performs Hadamard transformation
\begin{eqnarray}
H=\frac{1}{\sqrt{2}} \left( \begin{array}{c c} 1 & 1 \\
1 & -1
\end{array} \right)
\end{eqnarray}
on P$_i(B)$. If the state of [P$_i(A)$,P$_i(B)$,P$_i(C)$] is
$\ket{\psi_4}$, Bob and Charlie perform Hadamard transformation on
P$_i(B)$, P$_i(C)$, respectively. The Hadamard transformation is
crucial for the security of the scheme as we will see in the sequel.
After Bob and Charlie have done Hadamard transformation, they inform
Alice. Alice then selects randomly a sufficiently large subset of
particles from $A$ sequence, which we call $D$ sequence. Alice
generates a random bit string and encodes it on $D$ sequence. If
Alice's random bit is ``0''(``1''), she prepares a particle $a$ in
the state $\ket{+}=\frac{1}{\sqrt{2}}(\ket{0}+\ket{1})$
($\ket{-}=\frac{1}{\sqrt{2}}(\ket{0}-\ket{1})$) for each particle of
$D$ sequence. $D$ sequence is used to check eavesdropping, which we
call checking sequence. The remaining particles of $A$ sequence
forms $E$ sequence. $E$ sequence is used to encode Alice's secret
message, which we call message encoding sequence. Alice then encodes
her secret message on $E$ sequence. Similarly, if Alice's secret
message is ``0''(``1''), she prepares a particle $a$ in the state
$\ket{+}$ ($\ket{-}$) for each particle of $E$ sequence. Thus Alice
prepares $N$ particles for each particle of $A$ sequence, which we
call $a$ sequence [P$_1(a)$, P$_2(a)$,$\cdots$, P$_N(a)$].

(3) If the state of the particle P$_i(a)$ ($i=1,2,\cdots,N$) is
$\ket{+}$, then the state of the particle P$_i(a)$, P$_i(A)$,
P$_i(B)$, and P$_i(C)$ is
\begin{eqnarray}
\ket{\Phi_0}_{aABC}=\frac{1}{\sqrt{2}}(\ket{0}+\ket{1})_a\otimes\frac{1}{\sqrt{2}}(\ket{000}+\ket{111})_{ABC}.
\end{eqnarray}
where the subscript $a$ denotes the particle P$_i(a)$. If the
state of the particle P$_i(a)$ is $\ket{-}$, then the state of the
particle P$_i(a)$, P$_i(A)$, P$_i(B)$, and P$_i(C)$ is
\begin{eqnarray}
\ket{\Phi_1}_{aABC}=\frac{1}{\sqrt{2}}(\ket{0}-\ket{1})_a\otimes\frac{1}{\sqrt{2}}(\ket{000}+\ket{111})_{ABC}.
\end{eqnarray}

(4) Alice sends the particle P$_i(a)$, P$_i(A)$ through a CNOT
gate (P$_i(a)$ is the controller, P$_i(A)$ is the target). Then
$\ket{\Phi_0}_{aABC}$ becomes
\begin{eqnarray}
\ket{\Phi_0'}_{aABC}=\frac{1}{2}(\ket{0000}+\ket{1100}+\ket{0111}+\ket{1011})_{aABC},\nonumber\\
\end{eqnarray}
and $\ket{\Phi_1}_{aABC}$ is changed to
\begin{eqnarray}
\ket{\Phi_1'}_{aABC}=\frac{1}{2}(\ket{0000}-\ket{1100}+\ket{0111}-\ket{1011})_{aABC}.\nonumber\\
\end{eqnarray}

(5) Alice performs Hadamard transformation on the particle P$_i(a)$
and obtains
\begin{eqnarray}
\ket{\Phi_0''}_{aABC}&=&\frac{1}{2}[\ket{00}_{aA}\otimes\frac{1}{\sqrt{2}}(\ket{00}+\ket{11})_{BC}\nonumber\\
& &+\ket{10}_{aA}\otimes\frac{1}{\sqrt{2}}(\ket{00}-\ket{11})_{BC}\nonumber\\
& &+\ket{01}_{aA}\otimes\frac{1}{\sqrt{2}}(\ket{00}+\ket{11})_{BC}\nonumber\\
& &+\ket{11}_{aA}\otimes\frac{1}{\sqrt{2}}(\ket{11}-\ket{00})_{BC}].
\end{eqnarray}
or
\begin{eqnarray}
\ket{\Phi_1''}_{aABC}&=&\frac{1}{2}[\ket{00}_{aA}\otimes\frac{1}{\sqrt{2}}(\ket{00}-\ket{11})_{BC}\nonumber\\
& &+\ket{10}_{aA}\otimes\frac{1}{\sqrt{2}}(\ket{00}+\ket{11})_{BC}\nonumber\\
& &+\ket{01}_{aA}\otimes\frac{1}{\sqrt{2}}(\ket{11}-\ket{00})_{BC}\nonumber\\
& &+\ket{11}_{aA}\otimes\frac{1}{\sqrt{2}}(\ket{00}+\ket{11})_{BC}].
\end{eqnarray}

Noting that
\begin{eqnarray}
\frac{1}{\sqrt{2}}(\ket{00}+\ket{11})_{BC}=\frac{1}{\sqrt{2}}(\ket{++}+\ket{--})_{BC},
\end{eqnarray}
\begin{eqnarray}
\frac{1}{\sqrt{2}}(\ket{00}-\ket{11})_{BC}=\frac{1}{\sqrt{2}}(\ket{+-}+\ket{-+})_{BC},
\end{eqnarray}
we can also write
\begin{eqnarray}
\label{11}
\ket{\Phi_0''}_{aABC}&=&\frac{1}{2}[\ket{00}_{aA}\otimes\frac{1}{\sqrt{2}}(\ket{++}+\ket{--})_{BC}\nonumber\\
& &+\ket{10}_{aA}\otimes\frac{1}{\sqrt{2}}(\ket{+-}+\ket{-+})_{BC}\nonumber\\
& &+\ket{01}_{aA}\otimes\frac{1}{\sqrt{2}}(\ket{++}+\ket{--})_{BC}\nonumber\\
&
&-\ket{11}_{aA}\otimes\frac{1}{\sqrt{2}}(\ket{+-}+\ket{-+})_{BC}],\nonumber\\
\end{eqnarray}
\begin{eqnarray}
\label{12}
\ket{\Phi_1''}_{aABC}&=&\frac{1}{2}[\ket{00}_{aA}\otimes\frac{1}{\sqrt{2}}(\ket{+-}+\ket{-+})_{BC}\nonumber\\
& &+\ket{10}_{aA}\otimes\frac{1}{\sqrt{2}}(\ket{++}+\ket{++})_{BC}\nonumber\\
& &-\ket{01}_{aA}\otimes\frac{1}{\sqrt{2}}(\ket{+-}+\ket{-+})_{BC}\nonumber\\
&
&+\ket{11}_{aA}\otimes\frac{1}{\sqrt{2}}(\ket{++}+\ket{--})_{BC}].\nonumber\\
\end{eqnarray}

(6) Alice then measures the particle P$_i(a)$, P$_i(A)$ in the
$Z$-basis, $\{\ket{0}, \ket{1}\}$. Bob and Charlie measure the
particle P$_i(B)$, P$_i(C)$ in the $X$-basis, \{\ket{+},
\ket{-}\}, respectively. At this step, although Bob and Charlie
obtain their results of measurements, they cannot recover Alice's
secret message even if they collaborate, because they have no
information of Alice's result. We can draw the above conclusion
according to the equation \ref{11} and \ref{12}.

(7) Alice tells Bob and Charlie the order of $D$ sequence (checking
sequence). She randomly selects half of particles of $D$ sequence.
She lets Bob announce his results of measurements of the
corresponding particles of $B$ sequence firstly and then lets
Charlie announce his results of measurements of the corresponding
particles of $C$ sequence. For the other half of particles of $D$
sequence she lets Charlie firstly announce his results of
measurements and then does Bob. Alice judges whether her random bits
can be reconstructed correctly by combining Bob's and Charlie's
results. If the error rate is small, Alice can conclude that there
is no eavesdroppers in the line. Alice, Bob and Charlie continue to
perform the next step, otherwise they abort the communication.

(8) If Alice is certain that there is no eavesdropping, she
announces the results of measurements of $E$ sequence (message
encoding sequence). Thus Bob and Charlie can collaborate to recover
Alice's secret message, according to Alice's results, as illustrated
in Table 1.

\begin{table}[h]
\caption{The recovery of Alice's secret message }\label{Tab:one}
  \centering
    \begin{tabular}[b]{|c|c|c| c|} \hline
      Alice's result & Bob's result & Charlie's result & secret message\\ \hline
      \ 0 & \ket{+} & \ket{+} & 0\\ \hline
       \ 0 & \ket{+} & \ket{-} & 1\\ \hline
       \ 0 & \ket{-} & \ket{+} & 1\\ \hline
       \ 0& \ket{-} & \ket{-} & 0\\ \hline
      \ 1 & \ket{+} & \ket{+} & 1\\ \hline
       \ 1 & \ket{+} & \ket{-} & 0\\ \hline
       \ 1 & \ket{-} & \ket{+} & 0\\ \hline
       \ 1 & \ket{-} & \ket{-} & 1\\ \hline
    \end{tabular}
\end{table}
Suppose the results of Bob and Charlie are both $\ket{+}$. If
Alice's result of measurement of particle $a$ is ``0'' (``1''), they
then conclude that the Alice's secret message is ``0'' (``1'').

\section{Security of the Three-party QSS scheme}
\label{security}

So far we have proposed the three-party QSS scheme. We now discuss
the security of the present scheme. The crucial point is that the
randomly prepared four states $\ket{\psi_1}$, $\ket{\psi_2}$,
$\ket{\psi_3}$ and $\ket{\psi_4}$ do not allow an eavesdropper to
have a successful attack and the eavesdropper's attack will be
detected during the eavesdropping check.

We first consider the intercept-resend attack strategy. Suppose that
Bob is dishonest and he has managed to get a hold of Charlie's
particle as well as his own. We call the dishonest Bob, Bob*. Bob*
can only intercept $C$ sequence at the step 1 of the scheme and he
cannot make certain to which state does each of the intercepted
particles belong. $\ket{\psi_1}$, $\ket{\psi_2}$, $\ket{\psi_3}$ and
$\ket{\psi_4}$ can also be expressed as
\begin{eqnarray}
\ket{\psi_1}&=&\frac{1}{2}[\ket{+}_A(\ket{00}+\ket{11})_{BC}\nonumber\\
& &+\ket{-}_A(\ket{00}-\ket{11})_{BC}],
\end{eqnarray}
\begin{eqnarray}
\ket{\psi_2}&=&\frac{1}{2\sqrt{2}}\{[\ket{+}_A[(\ket{00}-\ket{11})+(\ket{01}+\ket{10})]_{BC}\nonumber\\
& &+\ket{-}_A[(\ket{00}+\ket{11})+(\ket{10}-\ket{01})]_{BC}\},
\end{eqnarray}
\begin{eqnarray}
\ket{\psi_3}&=&\frac{1}{2\sqrt{2}}\{[\ket{+}_A[(\ket{00}-\ket{11})+(\ket{01}+\ket{10})]_{BC}\nonumber\\
& &+\ket{-}_A[(\ket{00}+\ket{11})+(\ket{01}-\ket{10})]_{BC}\},
\end{eqnarray}
\begin{eqnarray}
\ket{\psi_4}&=&\frac{1}{2}[\ket{+}_A(\ket{00}+\ket{11})_{BC}\nonumber\\
& &+\ket{-}_A(\ket{01}+\ket{10})_{BC}].
\end{eqnarray}
Bob* measures P$_i(B)$, P$_i(C)$ in the Bell basis and then resends
P$_i(C)$ to Charlie. Suppose Bob* obtains
$\frac{1}{\sqrt{2}}(\ket{00}+\ket{11})$ after his Bell basis
measurement. If the original state of [P$_i(A)$, P$_i(B)$, P$_i(C)$]
is $\ket{\psi_2}$, then the state collapses to
$\ket{-}_A(\ket{00}+\ket{11})$. According to the scheme, the state
of [P$_i(a)$, P$_i(A)$, P$_i(B)$, P$_i(C)$] is changed to
$\ket{1-}_{aA}(\ket{+0}+\ket{-1})_{BC}$ or
$\ket{0-}_{aA}(\ket{+0}+\ket{-1})_{BC}$ at the step 6 of the scheme.
Without Alice's result, Bob* cannot have any information about
Alice's secret message. Moreover, Bob*'s eavesdropping will be
detected during the eavesdropping check. Because Charlie measures
his particle in the $X$ basis, Alice will find half of her secret
messages reconstructed by the results of Bob* and Charlie are
inconsistent with those of her. The error rate introduced by Bob*
will achieve 50\%. If the original state of [P$_i(A)$, P$_i(B)$,
P$_i(C)$] is $\ket{\psi_3}$, Bob*'s eavesdropping will also
introduce the error rate of 50\%. Similarly, suppose Bob* obtains
$\frac{1}{\sqrt{2}}(\ket{00}-\ket{11})$ after his Bell basis
measurement. If the original state of [P$_i(A)$, P$_i(B)$, P$_i(C)$]
is $\ket{\psi_2}$, then the state collapses to
$\ket{+}_A(\ket{00}-\ket{11})$. According to the scheme, the state
of [P$_i(a)$, P$_i(A)$, P$_i(B)$, P$_i(C)$] becomes
$\ket{0+}_{aA}(\ket{+0}-\ket{-1})_{BC}$ or
$\ket{1+}_{aA}(\ket{+0}+\ket{-1})_{BC}$ at the step 6 of the scheme.
During the eavesdropping check, Bob*'s eavesdropping will be
detected with probability 50\%. If the original state of [P$_i(A)$,
P$_i(B)$, P$_i(C)$] is $\ket{\psi_3}$, we can draw the same
conclusion. Similarly, if Bob* obtains
$\frac{1}{\sqrt{2}}(\ket{01}+\ket{10})$ or
$\frac{1}{\sqrt{2}}(\ket{01}-\ket{10})$, his eavesdropping will also
be detected during the eavesdropping check.

We then consider the collective attack strategy. If Bob* intercepts
Charlie's particle $C$ and uses it and his own ancillary particle
$B'$ in the state $\ket{0}$ to do a CNOT operation (the particle $C$
is the controller, Bob*'s ancillary particle, $B'$ is the target).
After that, Bob* resends the particle $C$ to Charlie. Bob* cannot
make certain to which state does the intercepted particle belong
since Alice prepared the three-particle states in $\ket{\psi_1}$,
$\ket{\psi_2}$, $\ket{\psi_3}$ or $\ket{\psi_4}$ randomly. Bob* can
performs CNOT operation or Hadamard plus CNOT operation in this
attack strategy. Suppose the original state of [P$_i(A)$, P$_i(B)$,
P$_i(C)$] is $\ket{\psi_1}$ or $\ket{\psi_2}$ and Bob* performs CNOT
operation on the intercepted particle and his ancillary particle. At
the step 5 of the scheme, the state of [P$_i(a)$, P$_i(A)$,
P$_i(B)$, P$_i(B')$, P$_i(C)$] will be
\begin{eqnarray}
\label{20}
\frac{1}{2}[\ket{00}_{aA}\otimes\frac{1}{\sqrt{2}}(\ket{000}+\ket{111})_{BB'C}\nonumber\\
+\ket{10}_{aA}\otimes\frac{1}{\sqrt{2}}(\ket{000}-\ket{111})_{BB'C}\nonumber\\
+\ket{01}_{aA}\otimes\frac{1}{\sqrt{2}}(\ket{000}+\ket{111})_{BB'C}\nonumber\\
+\ket{11}_{aA}\otimes\frac{1}{\sqrt{2}}(\ket{111}-\ket{000})_{BB'C}].
\end{eqnarray}
or
\begin{eqnarray}
\label{21}
\frac{1}{2}[\ket{00}_{aA}\otimes\frac{1}{\sqrt{2}}(\ket{000}-\ket{111})_{BB'C}\nonumber\\
+\ket{10}_{aA}\otimes\frac{1}{\sqrt{2}}(\ket{000}+\ket{111})_{BB'C}\nonumber\\
+\ket{01}_{aA}\otimes\frac{1}{\sqrt{2}}(\ket{111}-\ket{000})_{BB'C}\nonumber\\
+\ket{11}_{aA}\otimes\frac{1}{\sqrt{2}}(\ket{000}+\ket{111})_{BB'C}],
\end{eqnarray}
where the subscripts $B$ and $B'$ represent Bob*'s particle sent by
Alice and his ancillary particle, respectively. Note that
\begin{eqnarray}
\frac{1}{\sqrt{2}}(\ket{000}+\ket{111})_{BB'C}=\frac{1}{2}[(\ket{00}+\ket{11})_{BB'}\otimes\ket{+}_C\nonumber\\
+(\ket{00}-\ket{11})_{BB'}\otimes\ket{-}_C]\nonumber\\
\end{eqnarray}
and
\begin{eqnarray}
\frac{1}{\sqrt{2}}(\ket{000}-\ket{111})_{BB'C}=\frac{1}{2}[(\ket{00}-\ket{11})_{BB'}\otimes\ket{+}_C\nonumber\\
+(\ket{00}+\ket{11})_{BB'}\otimes\ket{-}_C].\nonumber\\
\end{eqnarray}
Bob* measures the particle $B, B'$, but he can only obtain
$\frac{1}{\sqrt{2}}(\ket{00}+\ket{11})_{BB'}$ or
$\frac{1}{\sqrt{2}}(\ket{00}-\ket{11})_{BB'}$, each with probability
1/2. Without Charlie's result, Bob* cannot obtain any information
about Alice's secret message. During the eavesdropping check, Bob*'s
eavesdropping will be detected since Alice randomly let Bob or
Charlie firstly announce their results of measurements. Suppose the
original state of [P$_i(A)$, P$_i(B)$, P$_i(C)$] is $\ket{\psi_3}$
or $\ket{\psi_4}$ and Bob* performs Hadamard plus CNOT operation on
the intercepted particle and his ancillary particle. Similarly, at
the step 5 of the scheme, the state of [P$_i(a)$, P$_i(A)$,
P$_i(B)$, P$_i(B')$, P$_i(C)$] can also be expressed as the equation
\ref{20} or \ref{21}. Thus Bob* cannot also obtain any information
about Alice's secret message without Charlie's result and his
eavesdropping will be detected by Alice easily.

\section{Multiparty QSS scheme}
\label{n-protocol}

We can easily generalize the three-party QSS scheme to a
$n$-party($n>3$) QSS one. Suppose that Alice want to send her secret
message to $n-1$ users. Alice prepare an ordered $N$ $n$-particle
states. Each of $n$-particle states is randomly in the state
\begin{eqnarray}
\ket{\Psi_1}=\frac{1}{\sqrt{2}}(\prod_{i=1}^n\ket{0}_{a_i}+\prod_{i=1}^n\ket{1}_{a_i}),
\end{eqnarray}
\begin{eqnarray}
\ket{\Psi_2}=\frac{1}{\sqrt{2}}(\ket{0+}_{a_1a_2}\prod_{i=3}^{n}\ket{0}_{a_i}+\ket{1-}{a_1a_2}\prod_{i=3}^{n}\ket{1}_{a_i}),
\end{eqnarray}
\qquad\qquad\qquad\qquad\qquad\qquad\vdots
\begin{eqnarray}
\ket{\Psi_{n}}=\frac{1}{\sqrt{2}}(\prod_{i=1}^{n-1}\ket{0}_{a_i}\ket{+}_{a_n}+\prod_{i=1}^{n-1}\ket{1}_{a_i}\ket{-}_{a_n}),
\end{eqnarray}
\begin{eqnarray}
\ket{\Psi_{n+1}}=\frac{1}{\sqrt{2}}(\ket{0}_{a_1}\prod_{i=2}^{n}\ket{+}_{a_i}+\ket{1}{a_1}\prod_{i=2}^{n}\ket{-}_{a_i}),
\end{eqnarray}
where $a_1$ denotes the sender, Alice's particle and $a_i$ denotes
the $i$-th user's particle ($i=1,2,\cdots,n$). The following steps
of the multiparty scheme is similar to those of the three-party
scheme. After transmitting the particles to $n-1$ users, Alice
announces publicly each of $n$-particle states she prepared. The
user, $a_i$ performs Hadamard transformation on his corresponding
particle according to Alice's information. Alice selects randomly a
sufficiently large subset of particles from her particle sequence,
which is formed checking sequence. Alice generates a random bit
string and encodes it on the checking sequence. The remaining
particles of Alice particle sequence forms message encoding
sequence, on which Alice encodes her secret message. Alice's random
bit string and secret message forms a $N$-bit string. Alice prepares
a particle, $a$ in the state $\ket{+}$ or $\ket{-}$ according to the
value of each bit of the $N$-bit string. She then performs a CNOT
operation on her own particles $a$ and $a_1$, where $a$ is the
controller, $a_1$ is the target. The state of the system becomes
\begin{eqnarray}
\frac{1}{\sqrt{2}}(\ket{00}_{aa_1}\otimes\prod_{i=2}^n\ket{0}_{a_i}+\ket{11}_{aa_1}\otimes\prod_{i=2}^n\ket{0}_{a_i}\nonumber\\
+\ket{01}_{aa_1}\otimes\prod_{i=2}^n\ket{1}_{a_i}+\ket{10}_{aa_1}\otimes\prod_{i=2}^n\ket{1}_{a_i})
\end{eqnarray}
or
\begin{eqnarray}
\frac{1}{\sqrt{2}}(\ket{00}_{aa_1}\otimes\prod_{i=2}^n\ket{0}_{a_i}-\ket{11}_{aa_1}\otimes\prod_{i=2}^n\ket{0}_{a_i}\nonumber\\
+\ket{01}_{aa_1}\otimes\prod_{i=2}^n\ket{1}_{a_i}-\ket{10}_{aa_1}\otimes\prod_{i=2}^n\ket{1}_{a_i}).
\end{eqnarray}
Alice performs Hadamard transformation on the particle $a$,
obtaining
\begin{eqnarray}
\frac{1}{\sqrt{2}}[\ket{0+}_{aa_1}\otimes\frac{1}{\sqrt{2}}(\prod_{i=2}^n\ket{0}_{a_i}+\prod_{i=2}^n\ket{1}_{a_i})\nonumber\\
+\ket{1-}_{aa_1}\otimes\frac{1}{\sqrt{2}}(\prod_{i=2}^n\ket{0}_{a_i}-\prod_{i=2}^n\ket{1}_{a_i})
\end{eqnarray}
or
\begin{eqnarray}
\frac{1}{\sqrt{2}}[\ket{0-}_{aa_1}\otimes\frac{1}{\sqrt{2}}(\prod_{i=2}^n\ket{0}_{a_i}-\prod_{i=2}^n\ket{1}_{a_i})\nonumber\\
+\ket{1+}_{aa_1}\otimes\frac{1}{\sqrt{2}}(\prod_{i=2}^n\ket{0}_{a_i}+\prod_{i=2}^n\ket{1}_{a_i})
\end{eqnarray}
Noting that
\begin{eqnarray}
\label{25}
\ket{\Omega_0}&=&\frac{1}{\sqrt{2}}(\prod_{i=2}^n\ket{0}_{a_i}+\prod_{i=2}^n\ket{1}_{a_i})\nonumber\\
&=&\frac{1}{2}[(\prod_{i=2}^{n-1}\ket{0}_{a_i}+\prod_{i=2}^{n-1}\ket{1}_{a_i})\otimes\ket{+}_{a_n}\nonumber\\
&
&+(\prod_{i=2}^{n-1}\ket{0}_{a_i}-\prod_{i=2}^{n-1}\ket{1}_{a_i})\otimes\ket{-}_{a_n}
\end{eqnarray}
and
\begin{eqnarray}
\label{26}
\ket{\Omega_1}&=&\frac{1}{\sqrt{2}}(\prod_{i=2}^n\ket{0}_{a_i}-\prod_{i=2}^n\ket{1}_{a_i})\nonumber\\
&=&\frac{1}{2}[(\prod_{i=2}^{n-1}\ket{0}_{a_i}-\prod_{i=2}^{n-1}\ket{1}_{a_i})\otimes\ket{+}_{a_n}\nonumber\\
&
&+(\prod_{i=2}^{n-1}\ket{0}_{a_i}+\prod_{i=2}^{n-1}\ket{1}_{a_i})\otimes\ket{-}_{a_n},
\end{eqnarray}
we can rewritten $\ket{\Omega_0}$, $\ket{\Omega_1}$ in the $X$-basis
by iterating the equation \ref{25}, \ref{26}. We then obtain
\begin{eqnarray}
\label{27}
\ket{\Omega_0}&=&\frac{1}{2^{n-3}\sqrt{2}}[\prod_{i=2}^n\ket{+}_{a_i}\nonumber\\
&
&+(\prod_{i=2}^{n-3}\ket{+}_{a_i})\otimes\ket{--}_{a_{n-1}a_{n}}+\cdots].
\end{eqnarray}
The right of the equation \ref{27} is the sum of $2^{n-2}$ terms,
each term is the permutation and combination of $\ket{+}, \ket{-}$
and the number of $\ket{-}$ in each term is even. we can also obtain
\begin{eqnarray}
\label{28}
\ket{\Omega_1}&=&\frac{1}{2^{n-3}\sqrt{2}}[(\prod_{i=2}^{n-1}\ket{+}_{a_i})\otimes\ket{-}_{a_n}\nonumber\\
&
&+(\prod_{i=2}^{n-3}\ket{+}_{a_i})\otimes\ket{-+}_{a_{n-1}a_{n}}+\cdots].
\end{eqnarray}
The right of the equation \ref{28} is also the sum of $2^{n-2}$
terms, each term is the permutation and combination of $\ket{+},
\ket{-}$, but the number of $\ket{-}$ in each term is odd. For
example
\begin{eqnarray}
\frac{1}{\sqrt{2}}(\ket{000}+\ket{111})&=&\frac{1}{2}(\ket{+++}+\ket{+--}\nonumber\\
& &+\ket{--+}+\ket{-+-}),
\end{eqnarray}
\begin{eqnarray}
\frac{1}{\sqrt{2}}(\ket{000}-\ket{111})_{BC}&=&\frac{1}{2}(\ket{++-}+\ket{+-+}\nonumber\\
& &+\ket{-++}+\ket{---}).
\end{eqnarray}

Alice then measures the particle $a, a_1$ in the $Z$-basis and the
receivers, $n-1$ users measure the particle $a_2,\cdots,a_n$ in the
$X$-basis, respectively. After doing these, Alice tells the users
the order of checking sequence and randomly lets them announce
firstly their results of measurements. Alice decides whether her
random bit string can be recovered correctly by collaborating the
$n-1$ user's results. If the error rate is small, Alice can conclude
that there is no eavesdroppers in the line. She then announces the
results of measurements of message encoding sequence. Thus the $n-1$
users can collaborate to recover Alice's message. The security
analysis of the multiparty QSS scheme is similar to that of the
three-party case, as described in Sec.\ref{security}.

\section{Summary}
\label{conclusion}

So far we have proposed an ($n, n$) threshold QSS scheme of secure
direct communication. Alice encodes her secret message into a given
state and sends it to the receivers directly using quantum channel.
Without Alice's result of measurement, the receivers cannot have any
information about Alice's secret message even if they collaborate.
Alice announces her result only if she is certain that there is no
eavesdropping in the line and only in such a way can the receivers
collaborate to recover Alice's secret message. By this token, the
present scheme is similar to QSDC protocol, different from the QSS
scheme whose task is essentially to allow a sender to establish a
shared key with the receivers. In our scheme, the states that Alice
prepared randomly are important that the eavesdropping of the
dishonest party or the eavesdropper, Eve can be detected easily. The
present scheme is efficient in that all the GHZ states used in the
scheme are used to generate shared secret messages except those
chosen for checking eavesdropper, the classical messages exchanged
between the communication parties only include the results of the
measurements, and the process of the scheme is also simple.



\begin{acknowledgments}
This work is supported by the National Natural Science Foundation of
China under Grant No. 60472032.
\end{acknowledgments}

%
%

%
%
\end{document}